%% file: aa25840-15.tex
\documentclass[online]{aa}
\usepackage[varg]{txfonts}
\RequirePackage{hyperref}
\hypersetup{
	colorlinks = true,
	linkcolor = blue,
	citecolor = blue,
	filecolor = blue,
	urlcolor = blue,
	pdfborder = {0 0 0}
}
\newcommand{\doiurl}[1]{\href{https://doi.org/#1}{#1}}
\newcounter{IonCS}
\DeclareRobustCommand{\ion}[2]{\setcounter{IonCS}{#2}\textup{#1\,\large{\scshape{\roman{IonCS}}}}}
\graphicspath{{./}{../Bilder/}}

\newcommand{\sect}[1]{Sect.\,\ref{S:#1}}

\newcommand{\fig}[1]{Fig.\,\ref{F:#1}}

\newcommand{\firstfig}[1]{\fig{#1}}
\newcommand{\tab}[1]{Table\,\ref{T:#1}}
\newcommand{\eqn}[1]{Eq.\,\eqnr{#1}}
\newcommand{\eqnf}[1]{Equation\,\eqnr{#1}}
\newcommand{\eqnr}[1]{(\ref{E:#1})}
\newcommand{\eqns}[2]{Eqs.\,\eqnr{#1} and \eqnr{#2}}

\newcommand{\tabl}[5]{\begin{table}\caption{#1}\label{T:#4}\centering\begin{tabular}{#2}\hline\hline#3\\\hline#5\\\hline\end{tabular}\end{table}}

\newcommand{\graphflex}[4][figure]{\begin{#1}#2\caption{#4}\label{F:#3}\end{#1}}
\newcommand{\graphsmall}[4][7cm]{\graphflex{\centering\resizebox{#1}{!}{\centering\includegraphics{#2.pdf}}}{#3}{#4}}
\newcommand{\graph}[3]{\graphflex{\resizebox{\hsize}{!}{\includegraphics{#1.pdf}}}{#2}{#3}}
\newcommand{\graphwidthflex}[6][figure*]{\graphflex[#1]{#5\includegraphics[width=#4]{#2.pdf}}{#3}{#6}}

\newcommand{\graphside}[4][12.8cm]{\graphwidthflex{#2}{#3}{#1}{\sidecaption}{#4}}

\newcommand{\eql}[1]{\begin{equation}#1\end{equation}}
\newcommand{\eqa}[1]{\begin{eqnarray}#1\end{eqnarray}}

\newcommand{\eqi}[1]{$#1$}

\DeclareRobustCommand*{\unit}[1]{~\ensuremath{\mathrm{#1}}}
\RequirePackage{color}
\definecolor{darkgreen}{rgb}{0,0.45,0}

\input{Journal.inc}

\begin{document}

%----------------------------------------------------------------------------
% A&A final format:
\setlength{\topmargin}{-20pt}
%----------------------------------------------------------------------------
\AANum{A86}
\yearCop{2016}
\doi{\doiurl{10.1051/0004-6361/201525840}}
\idline{A\&A 589, A86 (2016)}
\hypersetup{
	pdftitle = {Scaling laws of coronal loops compared to a 3D~MHD model of an active region},
	pdfauthor = {Ph.-A.~Bourdin, S.~Bingert, and H.~Peter},
	pdfkeywords = {Sun: corona -- magnetohydrodynamics (MHD) -- methods: numerical -- Sun: UV radiation},
	pdfsubject = {Astronomy \& Astrophysics}
}
%----------------------------------------------------------------------------

%=============================================================================
% TITLE
%=============================================================================
%
\title{Scaling laws of coronal loops compared to a 3D~MHD model of an active region}
\titlerunning{Scaling laws of coronal loops compared to an AR model}

\author{Ph.-A.~Bourdin\inst{1,2}, S.~Bingert\inst{3}, and H.~Peter\inst{2}}
\authorrunning{Ph.-A.~Bourdin et al.}
%\offprints{Ph.-A.~Bourdin}

\institute{%
Space Research Institute, Austrian Academy of Sciences, Schmiedlstr. 6, 8042 Graz, Austria
\\
\email{Philippe.Bourdin@oeaw.ac.at}
\and
Max-Planck-Institut f{\"u}r Sonnensystemforschung, Justus-von-Liebig-Weg 3, 37077 G{\"o}ttingen, Germany
\and
Gesellschaft f{\"u}r wissenschaftliche Datenverarbeitung, Am Fa{\ss}berg 11, 37077 G{\"o}ttingen, Germany
}

\date{Received 9 February 2015 / Accepted 16 March 2016}

\abstract%
%---CONETXT---(optional)-----------------------------------------------------
{%
The structure and heating of coronal loops have been investigated for decades.
Established scaling laws relate fundamental quantities like the loop apex temperature, pressure, length, and coronal heating.
}
%---AIMS----------------------------------------------------------------------
{%
We test these scaling laws against a large-scale 3D magneto-hydrodynamics (MHD) model of the solar corona, which became feasible with current high-performance computing.
}
%---METHODS-------------------------------------------------------------------
{%
We drove an active region simulation with photospheric observations and find strong similarities to the observed coronal loops in X-rays and extreme-ultraviolet (EUV) wavelength.
A 3D reconstruction of stereoscopic observations shows that our model loops have a realistic spatial structure.
We compared scaling laws to our model data extracted along an ensemble of field lines.
Finally, we fit a new scaling law that represents hot loops and also cooler structures, which was not possible before based only on observations.
}
%---RESULTS-------------------------------------------------------------------
{%
Our model data gives some support for scaling laws that were established for hot and EUV-emissive coronal loops.
For the Rosner-Tucker-Vaiana (RTV) scaling law we find an offset to our model data, which can be explained by 1D considerations of a static loop with a constant heat input and conduction.
With a fit to our model data we set up a new scaling law for the coronal heat input along magnetic field lines.
}
%--CONCLUIONS---(optional)----------------------------------------------------
{%
RTV-like scaling laws were fitted to hot loops and therefore do not predict well the coronal heat input for cooler structures that are barely observable.
The basic differences between 1D and self-consistent 3D modeling account for deviations between earlier scaling laws and ours.
We also conclude that a heating mechanism by MHD-turbulent dissipation within a braided flux tube would heat the corona stronger than is consistent with our model corona.
}
%-----------------------------------------------------------------------------
%----------------------------------------------------------------------------

\maketitle

%==============================================================================
\section{Introduction\label{S:energy.intro}}
%==============================================================================

The heating of the solar corona to millions of Kelvin is widely discussed in the literature \citep{Klimchuk:2006}.
On the one hand, wave heating (alternating current, AC) may transport large amounts of energy, but the exact nature of the dissipation mechanism able to reproduce realistic coronal structures remains unclear.
On the other hand, the field-line braiding mechanism proposed by \cite{Parker:1972} induces currents in the corona that are dissipated by Ohmic (direct current, DC) heating.
This is described well in magneto-hydrodynamics (MHD), and observationally driven computational models show an energy input together with a dissipation mechanism that creates loop structures within the corona that compare well to observations \citep{Bourdin+al:2013_overview,Bourdin+al:2014_coronal-loops,Bourdin+al:2015_energy-input}.

%------------------------------------------------------------------------------
\graphsmall{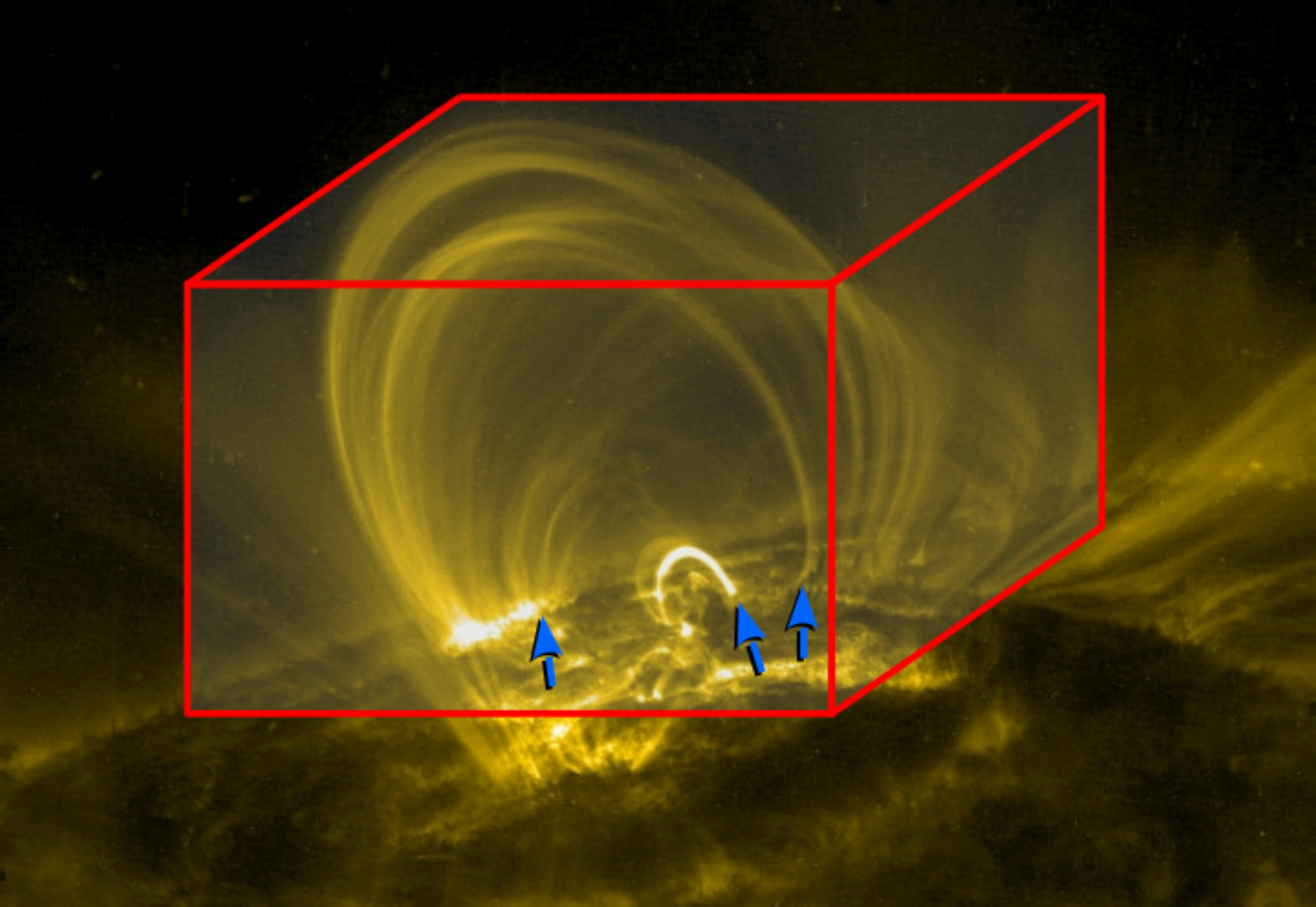}{TRACE}{
EUV emission of coronal loops above an active region (AR) observed by the {\sc TRACE} satellite \citep{Strong+al:1994}.
The footpoints of short, intermediate, and longer loops of varying curvature and intensity are indicated by blue arrows.
Typically, the short and most emissive loop is located in the core of an AR, while the longer loops may reach higher into the corona and may also connect to the periphery of the region.
The spatial extent of our 3D~MHD model is \eqi{235 \times 235 \times 156\unit{Mm^3}} and is roughly indicated by the red box.
Image credit: NASA/LMSAL.}
%------------------------------------------------------------------------------

Later, \cite{Parker:1988} extended the field-line braiding idea towards the actual dissipation mechanism, where so-called nanoflares are proposed.
Nanoflares are short-lived strong heating events after fast and small-scale magnetic reconnection due to the braiding of coronal field lines.
In reality, the advection of magnetic field lines in the photosphere is too slow to create a substantial braid in the corona because the Alfv{\'e}n speed is high enough to immediately propagate all disturbances in the magnetic field into the corona, so that it might be preferable to consider the process as a ``quasi-static reconfiguration and subsequent magnetic energy dissipation'' \citep{Bourdin+al:2015_energy-input}.
This process can still yield a strong particle acceleration in the corona, e.g., to non-thermal energies for electrons on the order of several \unit{MeV}, as demonstrated in \cite{Threlfall+Bourdin:2016_particle-acceleration}.

%------------------------------------------------------------------------------
\graphside{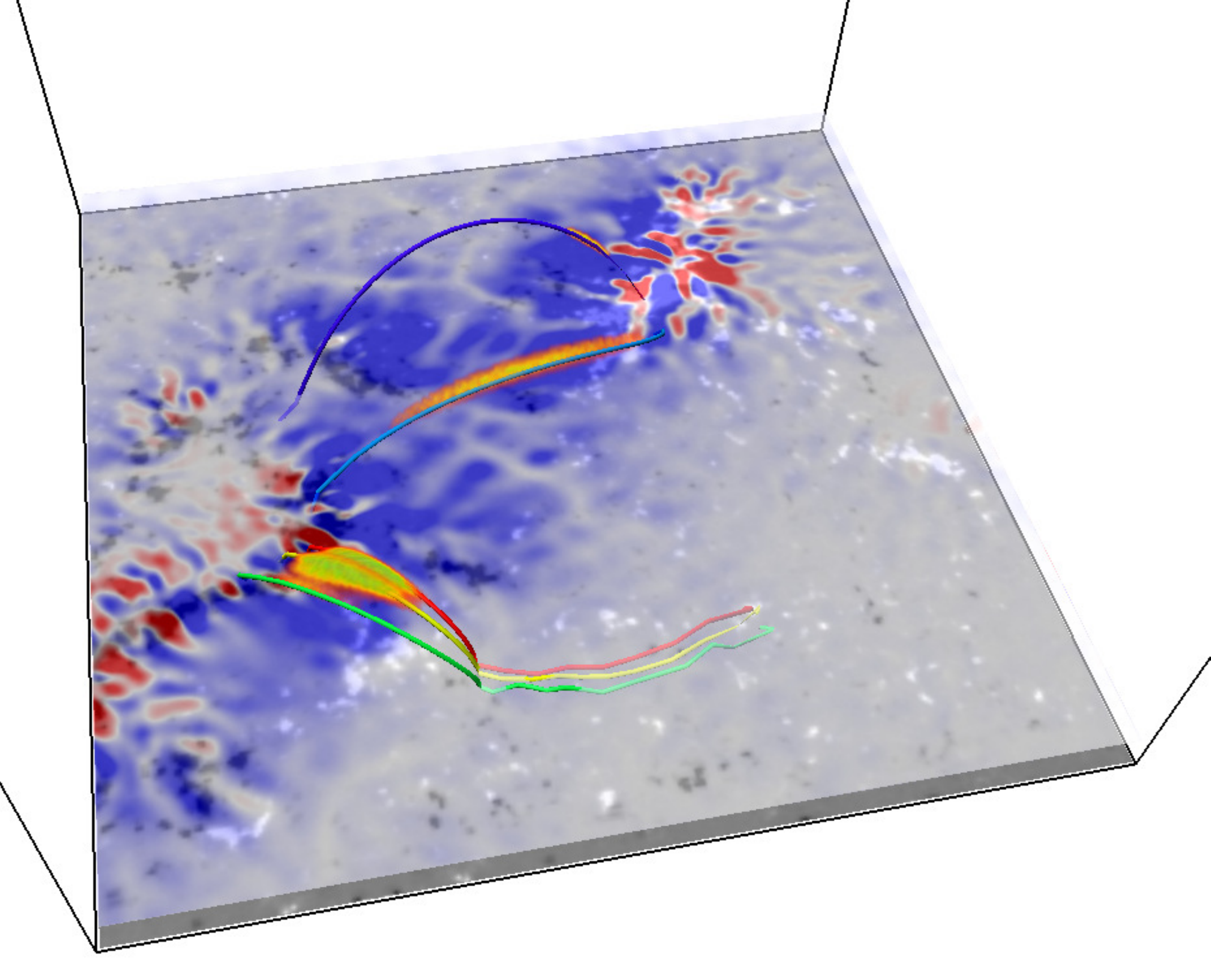}{AR.overview}{
Three-dimensional visualization of the 3D~MHD model with the line-of-sight magnetogram of the AR at the base layer.
The magnetic flux density is coded in grayscale with a saturation level of \eqi{\pm\,300\unit{Gauss}} (black and white).
At a height of 3\unit{Mm} we overlay (half-transparent) the vertical component of the Poynting flux with a saturation level of \eqi{\pm\,50\,000\unit{W/m^2}} color-coded in blue-transparent-red, where the zero level is transparent.
The EUV emission of the \ion{Fe}{15} line extracted from the model is color-coded in orange-red-transparent and the field lines crossing the respective emission maxima are drawn as tubes in red (SL\,1), yellow (SL\,2), green (SL\,3), light blue (CL\,1), and dark blue (CL\,2).
The black box boundaries indicate the central \eqi{117 \times 117\unit{Mm^2}} AR core area.}
%------------------------------------------------------------------------------

A different approach to understanding and describing the heating of the corona is to statistically derive scaling laws from observed coronal loops that relate global loop properties such as their maximum temperature, their length, their density, or the magnetic field at their footpoints with the heating rate \citep{Rosner+al:1978,vanBallegooijen:2011}, irrespective of exactly how this heating is produced.
Such scaling laws have also been used to drive coronal models \citep[e.g.,][]{Lionello+al:2005} and to test the observable consequences of different scaling laws on the corona \citep{vanWettum+al:2013}.
With observations of an active region and a numerical experiment that matches observations \citep{Bourdin+al:2013_overview,Bourdin+al:2014_coronal-loops,Bourdin+al:2015_energy-input}, we are now able to test theoretical scaling laws (that relate global coronal loop parameters with their heating) for consistency with our model corona.
For that, we extract different physical quantities along an ensemble of model field lines and check the scaling-law predictions.
The temperatures of the hot extreme-ultraviolet (EUV)-bright loops in our model compare well to empirical scaling laws \citep{Rosner+al:1978}.
Loops that span towards the periphery are usually less emissive (see rightmost marked footpoint in \firstfig{TRACE}) because they are rooted with at least one footpoint in areas with a lower magnetic energy input to the corona.
Hence, these loops are cooler, which -- at first look -- speaks against the results of existing empirical scaling laws, like found by \cite{Rosner+al:1978}.
For the coronal heating and the resulting temperature we fit a new scaling law and compare this with other MHD model results \citep{Rappazzo+al:2008,vanBallegooijen:2011}.

%------------------------------------------------------------------------------
\section{Coronal model and field-line ensemble\label{S:coronal.model}}
%------------------------------------------------------------------------------

The model is driven only from the lower boundary with an observed line-of-sight magnetogram (see \firstfig{AR.overview}).
The magnetic energy input in the form of Poynting flux is through a photospheric motion driver that contains the large-scale motions of the magnetic patches as deduced from the observed timeseries of magnetograms together with a photospheric motion driver, as we describe in more detail in \cite{Bourdin+al:2013_overview}.
As an initial condition we used a potential field extrapolation, but then propagated the magnetic field fully self-consistently by the MHD equations.
We started the simulation smoothly to avoid switch-on effects \citep[see][]{Bourdin:2014_switch-on} and to compensate for a lack of coronal heating until the magnetic disturbances from the photosphere have propagated with the Alfv{\'e}n speed into the corona, as we describe in \cite{Bourdin+al:2014_coronal-loops}.
For the simulation we used the Pencil Code \citep{Brandenburg+Dobler:2002}\footnote{\url{http://Pencil-Code.Nordita.org/}}.

We selected for our analysis the same set of field lines as in our previous publication \citep{Bourdin+al:2015_energy-input}.
These field lines were selected from seed points randomly distributed in a volume over the active region (AR) core.
An additional set of seed points was distributed on a vertical plane that cuts through most of the closed-field lines between the two main polarities along solar-Y direction.
The distribution of seed points is weighted by the magnetic flux density.
We selected field lines that a) are closed and do not connect to the upper box boundary; b) have a length of 18\unit{Mm} to 150\unit{Mm}; and c) reach a minimum temperature of at least 75\,000\unit{K} before reaching a height of 18\unit{Mm}.
 We also made sure that the field lines crossing the EUV-intensity maximum of the most prominent loops \citep[SL\,1-3 and CL\,1+2; see][]{Bourdin+al:2013_overview} were included in this ensemble of about 67\,000 field lines, as well as approximately 200 field lines neighboring these loops.

%------------------------------------------------------------------------------
\section{Scaling laws for coronal loop properties\label{S:scaling.laws}}
%------------------------------------------------------------------------------
We have good indications for relations of the coronal Ohmic heating to the field-line length, the footpoint magnetic flux density, and hence also the vertical net Poynting flux, \citep{Bourdin+al:2015_energy-input}.
We now want to compare our model field lines with theoretical scaling laws relating plasma parameters (such as density, temperature, or pressure) of the field line to the energy input and field-line length.
These scaling laws are also partly derived from observations \citep{Rosner+al:1978}.
We refer to these scaling laws as RTV, named after the authors.
\eqa{T &=& c_T \cdot {(p L)}^{1/3} \label{E:RTV_T} ,\\
H &=& c_H \cdot {p}^{7/6} L^{-5/6} \label{E:RTV_H} ,}
for the temperature \eqi{T} and the heating rate \eqi{H} along a hot coronal loop.
Here, the heating rate \eqi{H(s)} is assumed to be constant along a loop, irrespective of the actual mechanism that delivers the energy.

The RTV equations can be rewritten as
\eqa{p &=& c_H^{-6/7} \cdot H^{6/7} L^{5/7} \label{E:RTV_p_H} ,\\
T &=& c_T \cdot c_H^{-2/7} \cdot H^{2/7} L^{4/7} \label{E:RTV_T_H} .}

With the ideal gas law \eqi{p=2 n_{\rm{e}} k_{\rm{B}} T} we find
\eql{n_{\rm{e}} = \frac{1}{2 k_{\rm{B}}} c_H^{-4/7} c_T^{-1} \cdot H^{4/7} L^{1/7} \label{E:RTV_n_H}}
for the electron number density \eqi{n_{\rm{e}}}.

In our model we have a variable Ohmic heating \eqi{H(s)} deposited along the field lines, so it makes more sense to use the integrated heating
\eql{F_H = \int {H(s) \cdot ds} .}
For a roughly constant \eqi{H}, this simplifies to \eqi{F = H \cdot L}.
Under this assumption \eqns{RTV_T_H}{RTV_n_H} become
\eqa{T &=& c_T \cdot c_H^{-2/7} \cdot {F_H}^{2/7} L^{2/7} \label{E:RTV_T_F} ,\\
n_{\rm{e}} &=& \frac{1}{2 k_{\rm{B}}} c_H^{-4/7} c_T^{-1} \cdot {F_H}^{4/7} L^{-3/7} \label{E:RTV_n_F} .}

\eqnf{RTV_T_H} can be rewritten as a relation of the loop heating
\eql{H = c_H \cdot c_T^{-7/2} \cdot T^{7/2} L^{-2} \label{E:RTV_H_F} .}

%------------------------------------------------------------------------------
\section{Scaling laws in a 3D model\label{S:model.scaling.laws}}
%------------------------------------------------------------------------------

%------------------------------------------------------------------------------
\subsection{Heating and loop length\label{S:heating.length}}
%------------------------------------------------------------------------------

%------------------------------------------------------------------------------
\graph{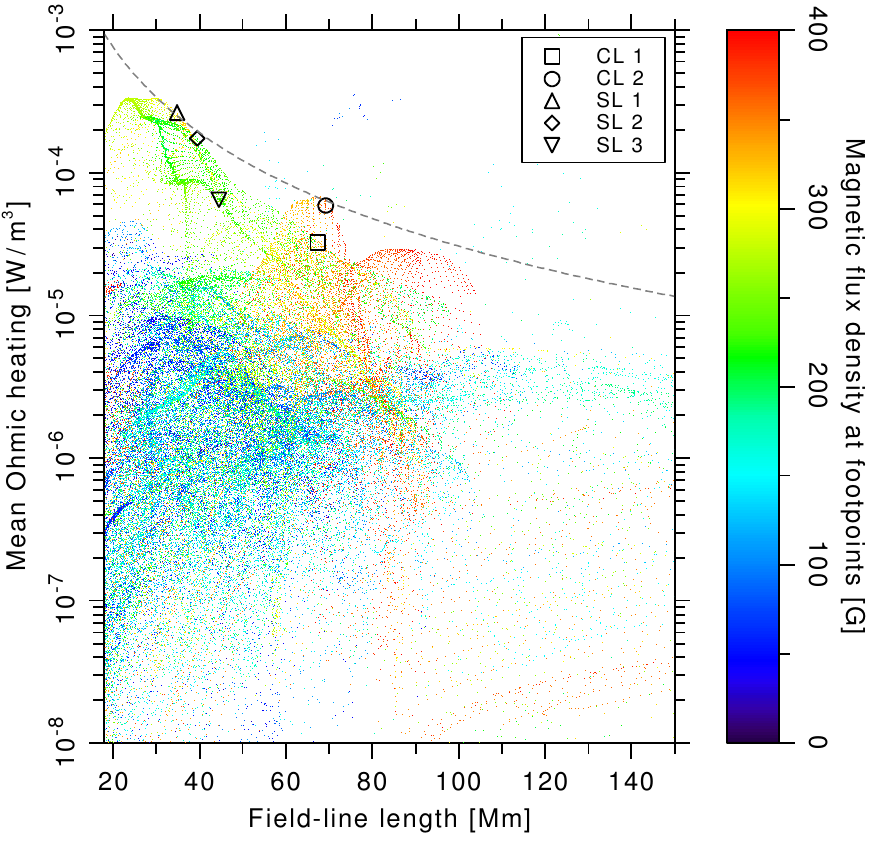}{L_HR_ohm_total}{
Field-line length versus volumetric Ohmic heating averaged along the field line together with the magnetic flux density at the field-line footpoints (color coded) at a height of 300\unit{km}.
The gray dashed curve represents the RTV scaling law following \eqn{RTV_H_F} for a mean Ohmic heating \eqi{\langle H \rangle = F_H / L} of a sample coronal loop (see \sect{heating.length}).}
%------------------------------------------------------------------------------

%------------------------------------------------------------------------------
\graph{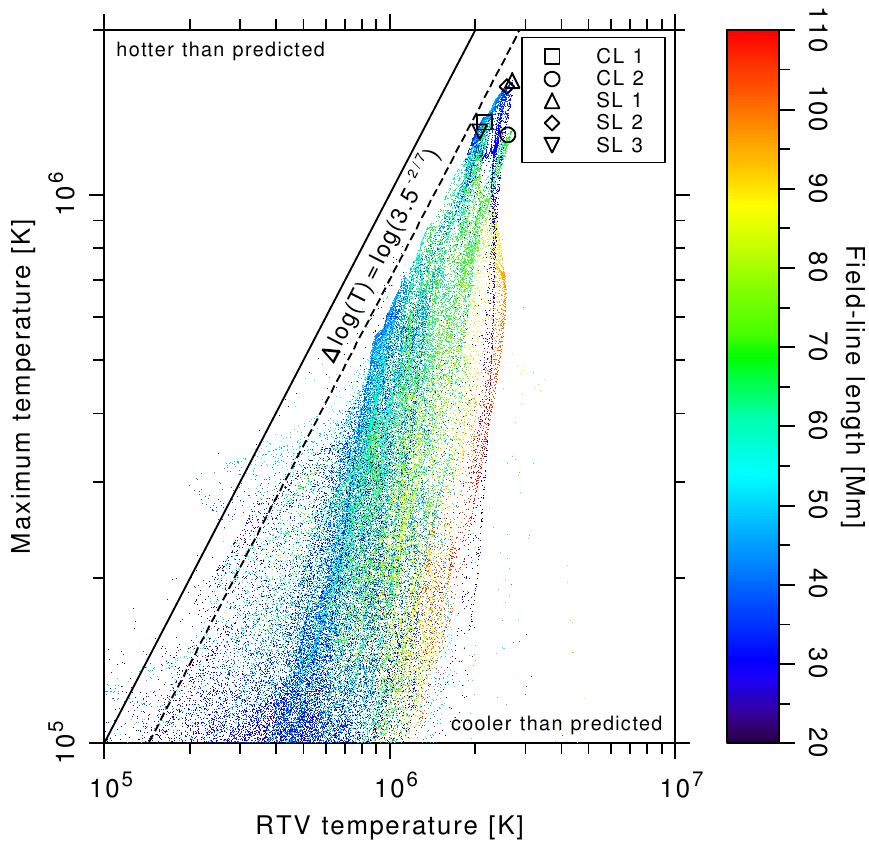}{Temp_RTV_max_Temp_max}{
Maximum temperature along field lines versus temperature predicted by the RTV scaling law as given in \eqn{RTV_T_F}.
The solid line indicates the equality to the RTV scaling law, while the dashed line indicates a correction factor of \eqi{3.5} (see \sect{loop.temperature}).}
%------------------------------------------------------------------------------

First, we plot in \firstfig{L_HR_ohm_total} the field-line length versus the average of the volumetric Ohmic heating; both quantities represent only the coronal part, as defined in \sect{coronal.model}.
We find a broad distribution, where field lines with footpoint flux densities of 300-400\unit{G} appear practically everywhere, but there is a concentration in the upper center around 75-100\unit{Mm} field-line length that shows significantly high heating rates.
As we go to shorter field lines, we see that the footpoint flux density decreases to values around 200\unit{G} for field-line lengths of 20\unit{Mm}.
At the same time the Ohmic heating rises, following the trend of the overplotted RTV scaling law (gray dashed line) that we get from sample parameters of a hot coronal loop with a maximum temperature of \eqi{T_{\rm{max}}=1.6\unit{MK}}.
The three hottest coronal loops (SL\,1+2 and CL\,1) closely follow this RTV scaling law.
The same relation holds also between both warm loops (SL\,3 and CL\,2).

As we already demonstrated \citep[see Fig.\,10 in][]{Bourdin+al:2015_energy-input}, field lines with footpoint flux densities below a critical limit of 200\unit{G} do not show particularly high Ohmic heating.
Nonetheless, they can span short and long distances into the corona.
For field lines longer than 150\unit{Mm} with flux densities of about 150\unit{G} (turquoise), we still find some that reach towards the scaled properties of much shorter warm coronal field lines (cf. \fig{L_HR_ohm_total}).
If this scenario is persistent, more warm loops of lengths above 150\unit{Mm} will form in our model high up in the corona.
In addition, there is also a large population of field lines with low footpoint flux densities that are simply not sufficiently heated and that are currently cooling down.
This is consistent with coronal observations where very few coronal field lines are actually seen as loops that are bright in EUV or X-ray emission lines and the majority are not.
Therefore, our finding that \eqi{H \sim L^{-2}} provides an upper limit for the heat input in our model is consistent with the general trend from the RTV scaling law.
Still, we find quite a large scatter of the RTV predictions for field lines along which the heating is less strong and that are consequently loaded with cooler plasma.

%------------------------------------------------------------------------------
\subsection{Loop temperature\label{S:loop.temperature}}
%------------------------------------------------------------------------------

The RTV scaling law itself can also be tested against the model directly.
We do this by comparing the maximum temperatures of the coronal loop, once taken directly from the model and once calculated from the model properties, i.e., the energy input into a structure and its field-line length using the RTV scaling law (\eqn{RTV_T_F}).
Because RTV assumes a constant heating rate, here we use the average heating rate \eqi{\langle H \rangle = F_H / L} for comparison.
In \firstfig{Temp_RTV_max_Temp_max} we present the distribution of the model values in relation to the RTV scaling law where the black solid line indicates the equality of both quantities.

In hydrostatic equilibrium and with a prescribed heating, the loop-top temperature \eqi{T_{\rm{max}}} can be derived for a 1D loop model in thermal equilibrium and with a static energy balance including radiative losses and heat conduction.
As a result, the RTV heating (\eqn{RTV_H}) is derived with a constant correction factor of \eqi{3.5} \citep{Priest:1982}.
This correction factor corresponds to a constant shift in logarithmic coordinates (dashed line in \fig{Temp_RTV_max_Temp_max}).
The slope of both lines reflects an exponent of \eqi{2/7}.

We find a good match in the scaling for the hot end of the distribution, starting with temperatures of 0.5\unit{MK} and up to our hottest model loops (symbols) at around 1.7\unit{MK} (see \fig{Temp_RTV_max_Temp_max}).
Also here, a large population of field lines is cooler than predicted, which is not a contradiction, because these cool field lines are not covered by the RTV scaling laws that were laid out to match only hot (EUV-emissive) loops.

%------------------------------------------------------------------------------
\section{Parameterization of Ohmic heating\label{S:parametrization}}
%------------------------------------------------------------------------------

%------------------------------------------------------------------------------
\subsection{van Ballegooijen scaling law\label{S:Ballegooijen.scaling}}
%------------------------------------------------------------------------------

%------------------------------------------------------------------------------
\graph{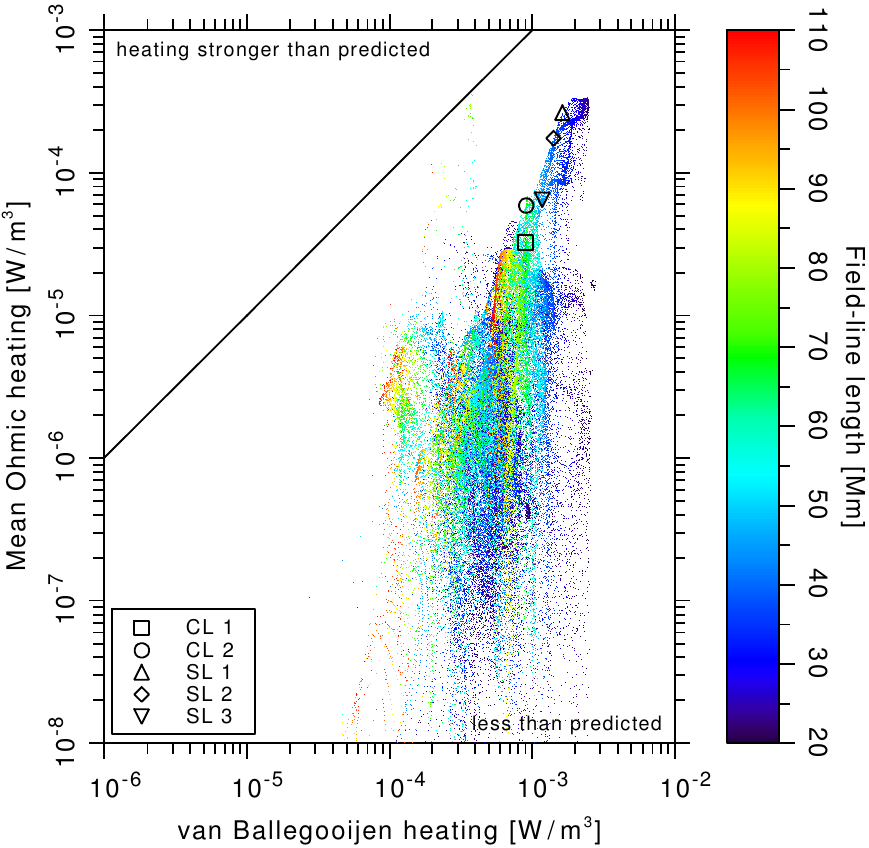}{HR_Ballegooijen_HR_ohm_total}{
Mean Ohmic heating as predicted from the \cite{vanBallegooijen:2011} scaling law versus the mean Ohmic heating along the model field lines.
The black line indicates the equality between the predicted and the model values
(see \sect{Ballegooijen.scaling}).}
%------------------------------------------------------------------------------

\vspace{6pt}

We want to investigate additional possible scaling laws for the Ohmic heating rate.
For instance, \cite{vanBallegooijen:2011} have conducted box model simulations of individual coronal loop structures (flux tubes) using a stratified atmosphere.
In these models, driving motions are applied on the two opposite ends of the cuboid box, which encompasses the non-curved straight loop.
Alfv{\'e}n waves and MHD turbulences heat the coronal part of the loop.
From a parameter study, the following relation to the loop length, the magnetic flux density at the photospheric boundary, and the driving motions is inferred
\eqa{Q_{\rm{cor}} &=& 2.9 \cdot 10^{-3} \left({{0.45}+\frac{33\unit{s}}{\tau_0}}\right) \left({\frac{B_{\rm{cor}}}{50\unit{G}}}\right)^{\beta} \nonumber\\
&& \cdot \left({\frac{L}{50\unit{Mm}}}\right)^{\gamma} \left({\frac{v_{\rm{rms}}}{1.48\unit{km/s}}}\right)^{\delta}\unit{\left[\frac{W}{m^3}\right]}\label{E:Ballegooijen},}
where \eqi{\tau_0 = 60-200\unit{s}} is the correlation time of the driving motions.
The average velocity (root-mean-squared) of the driving motions in our model is \eqi{v_{\rm{rms}} = 1.242\unit{km/s}}.
We use the exponents for the relation between coronal heating and magnetic field strength (\eqi{\beta}), the driving velocity relation (\eqi{\delta}), and the field-line length relation (\eqi{\gamma}) as given in \cite{vanBallegooijen:2011} (see also \tab{exponents}).
In their scaling law the driving motions \eqi{v_{\rm{rms}}} were of larger amplitude (1.48\unit{km/s}) and hence we expect a 25\unit{\%} lower heating rate prediction owing to the 16\unit{\%} weaker driving motions in our model.

In \firstfig{HR_Ballegooijen_HR_ohm_total} we compare the parameters of our model field lines with the prediction based on the \cite{vanBallegooijen:2011} scaling law for the coronal heating.
We find that practically all coronal field lines have a significantly lower heating rate than predicted, which is not explicable by our weaker driving motions.
The short and hot AR core loops (SL\,1+2) have the highest predicted values for the coronal heating, while the longer loops (CL\,1+2) are heated about ten times less.
These strongest heated field lines form a population (from CL\,1 to SL\,1, above \eqi{3 \cdot 10^{-5}\unit{W/m^3}}) that indicates a slope roughly twice as steep as the \cite{vanBallegooijen:2011} scaling law would predict.

The length relation in \eqn{Ballegooijen} seems to be partly responsible for the distribution of the data points within this most heated population.
As one would expect from the inverse dependency on \eqi{L}, the shorter field lines (blue) have a higher predicted coronal heating than the longer field lines (turquoise).
This trend even continues to the longest coronal field lines (yellow and red) that are located below the long hot coronal loop CL\,1.

For a coronal Ohmic heating below \eqi{10^{-5}\unit{W/m^3}} we find a strong variation of the field-line parameters that does not allow conclusions on the slope of possible scaling laws.
A dependency on the field-line length can still be seen because the long (red) and the short (blue) field lines are roughly ordered from left to right.
Overall, the match of the prediction to our model parameters is not good.

%------------------------------------------------------------------------------
\subsection{Rappazzo scaling law\label{S:Rappazzo.scaling}}
%------------------------------------------------------------------------------

%------------------------------------------------------------------------------
\graph{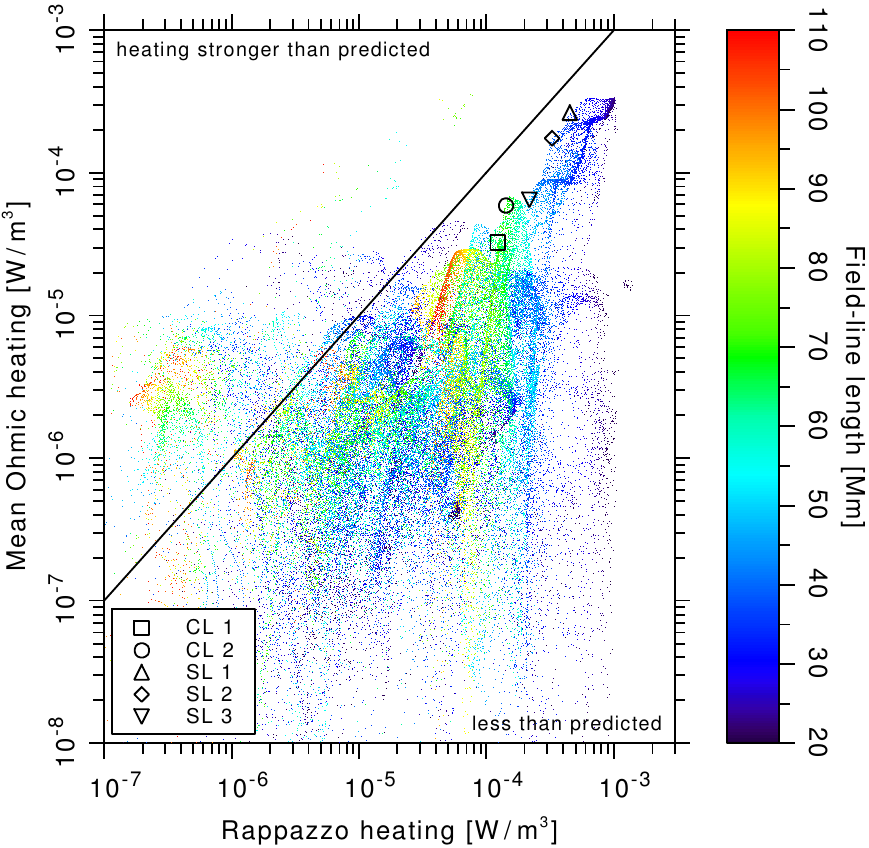}{HR_Rappazzo_HR_ohm_total}{
Same as \fig{HR_Ballegooijen_HR_ohm_total}, but for the scaling law and the exponents from \cite{Rappazzo+al:2008}.}
%------------------------------------------------------------------------------

Based on MHD turbulence models that resemble the field-line braiding mechanism, \cite{Rappazzo+al:2008} have found a different dependency of the Ohmic heating on the magnetic flux density and the loop length from \cite{vanBallegooijen:2011}.
The exponents \eqi{\beta}, \eqi{\delta}, and \eqi{\gamma} (see \tab{exponents}), together with the additional dependency \eqi{\omega = 0.125} on the mean field-line particle number density \eqi{n_\rho}, describe a scaling law of the form
\eqa{Q_{\rm{turb}} &=& 2.9 \cdot 10^{-3} \frac{120\unit{s}}{\tau_A+60\unit{s}} \left({\frac{n_\rho}{10^{15}}}\right)^{\omega} \left({\frac{B_{\rm{cor}}}{50\unit{G}}}\right)^{\beta} \nonumber\\
&& \cdot \left({\frac{L}{50\unit{Mm}}}\right)^{\gamma} \left({\frac{v_{\rm{rms}}}{1.48\unit{km/s}}}\right)^{\delta}\unit{\left[\frac{W}{m^3}\right]} \label{E:Rappazzo}.}
In \cite{vanWettum+al:2013} this form prescribes one of several tested heating parametrizations for a coronal MHD model.

In contrast to the \cite{vanBallegooijen:2011} scaling law, \cite{Rappazzo+al:2008} use the Alfv{\'e}n crossing time \eqi{\tau_A} as a time scale and not the correlation time scale \eqi{\tau_0} of the driving motions.
Because the value range of \eqi{\tau_A} reaches 2000\unit{s}, we need to adapt the \eqi{\tau} term of the scaling law so that it represents not only the value range between 60--200\unit{s} \citep{vanBallegooijen:2011}, but also has a continuous effect on larger \eqi{\tau} values.
 In our case loops have a length of about 75\unit{Mm}, which implies \eqi{\tau_A = 750\unit{s}} for average Alfv{\'e}n velocities of about 100\unit{km/s}.

In \firstfig{HR_Rappazzo_HR_ohm_total} we see that the \cite{Rappazzo+al:2008} exponents describe our model data better than the \cite{vanBallegooijen:2011} scaling law.
The slope of the strongly heated field-line population (extending towards the upper right in \fig{HR_Rappazzo_HR_ohm_total}) fits the equality line (black) more closely.
Furthermore, the discrepancy between the prediction and the model values is much smaller than it is in \fig{HR_Ballegooijen_HR_ohm_total}.
This shows that the relation between the Ohmic heating to the magnetic field strength is weaker, and the relation to the field-line length is stronger in our model than predicted by the \cite{Rappazzo+al:2008} scaling law.

%------------------------------------------------------------------------------
\subsection{Fitting a new scaling law\label{S:fitted.scaling}}
%------------------------------------------------------------------------------

We can of course also try to find better exponents for a purported scaling law of the form
\eqa{Q_{\rm{Ohm}} &=& 2.9 \cdot 10^{-3} \frac{120\unit{s}}{\tau_A+60\unit{s}} \left({\frac{n_\rho}{10^{15}}}\right)^{\omega} \left({\frac{B_{\rm{cor}}}{50\unit{G}}}\right)^{\beta} \nonumber\\
&& \cdot \left({\frac{L}{50\unit{Mm}}}\right)^{\gamma} \left({\frac{v_{\rm{rms}}}{1.48\unit{km/s}}}\right)^{\delta}\unit{\left[\frac{W}{m^3}\right]} \label{E:modified.scaling}.}
For this we use a Levenberg-Marquardt (LM) optimization to adjust the exponents \eqi{\beta}, \eqi{\delta}, and \eqi{\gamma} in \eqn{modified.scaling}.
We keep the dependency on the mean particle number density fixed because there is no correlation to fit to in our model data (see Fig.\, 6 in \cite{Bourdin+al:2014_coronal-loops}).
We use the same exponents as in \sect{Rappazzo.scaling}, except that in our model the Alfv{\'e}n crossing time \eqi{\tau_A} through a loop is on the order of 40 minutes.

We exclude the data points with a heating rate lower than the threshold value of \eqi{5 \cdot 10^{-6}\unit{W/m^3}} because there is practically no structure in this data and including it would result in a false impression of accuracy by underestimated error intervals of the fitted exponents.
For the fitting, we estimate the error of the model data to a constant value of 25\unit{\%} of the maximum heating rate.
This reflects the large scatter we see in the data and actually gives less weight to the weakly heated field lines.

%------------------------------------------------------------------------------
\graph{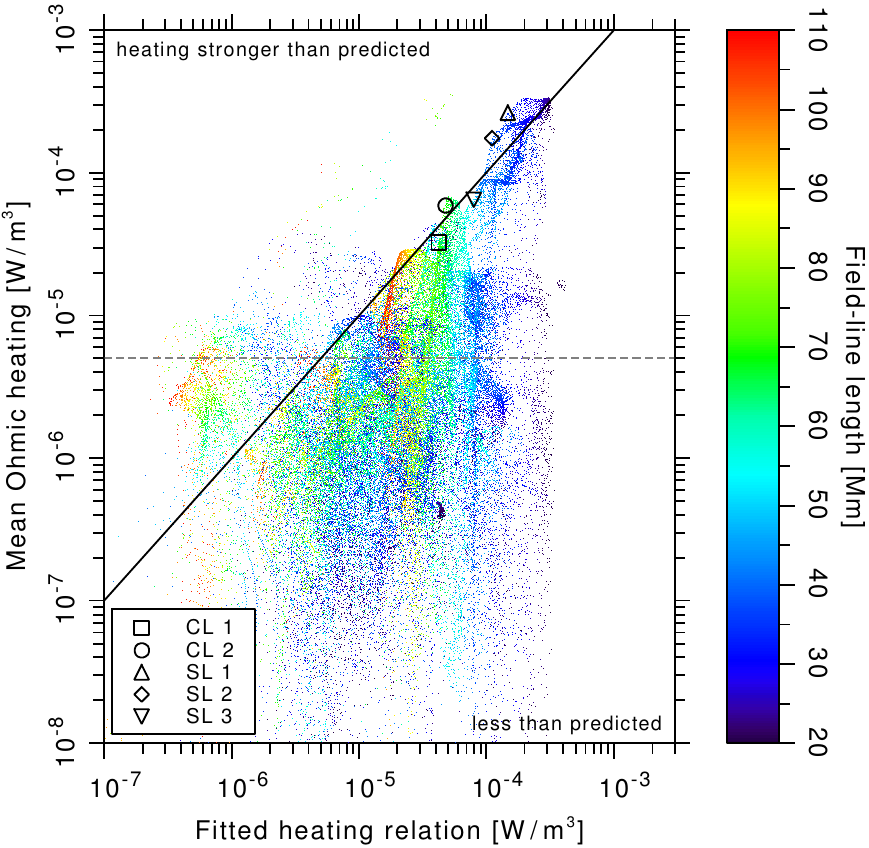}{HR_modified_HR_ohm_total}{
Volumetric Ohmic heating as predicted from our fitted scaling law versus the mean Ohmic heating along the model field lines.
On the black line the prediction and the model values are equal.
The gray dashed line indicates the cutoff value we use for fitting the data points (see \sect{fitted.scaling}).}
%------------------------------------------------------------------------------

We fixed the parameter \eqi{\tau_A} during the optimization because it acts on the same degree of freedom (DOF) as \eqi{\delta} does, namely shifting the whole data towards higher or lower heating rates.
Freeing two parameters for one DOF would result in an underestimated \eqi{\chi} and overestimated errors for the fitted parameters.
We keep \eqi{\omega = 0.125}, which follows from setting \eqi{\alpha = 2} in the \cite{Rappazzo+al:2008} scaling law.

From the LM optimization of the 20\,000 remaining field lines that surpass the heating rate threshold, we get the exponents with a reduced \eqi{\chi} of 2.45 per DOF (see \tab{exponents}).

A \eqi{\chi} larger than \eqi{1} indicates missing degrees of freedom for the fit or, alternatively, an error estimate that is too small, but our error estimate is already quite large.
A reduced \eqi{\chi} on the order of \eqi{2}, as we find here, would therefore indicate that there should be roughly twice as many DOFs as we give to the fitting procedure.
In this case, we can safely state that there are more relations to consider in order to predict the coronal Ohmic heating than just the field-line length, the density, and the magnetic field strength.
For example, it could make a difference if the used footpoint locations of a field line are relatively low or higher up in the atmosphere because this already changes the net Poynting flux into the field line in average \citep[see Fig.\,2 in][]{Bourdin+al:2015_energy-input}.
Also, the highly varying net vertical Poynting flux at the field-line footpoints has an influence irrespective of the magnetic field strength.

\tabl{Exponents for different coronal heating scaling laws as used in \eqns{Ballegooijen}{Rappazzo} together with the fitting results from \sect{fitted.scaling}.}{c c c c}{Parameter & van Ballegooijen & Rappazzo & this work \\ & & \eqi{(\alpha = 2)} & }{exponents}{
\eqi{\beta} & \eqi{+0.55} & \eqi{+1.75} & \eqi{+1.25 \pm 0.10} \\
\eqi{\gamma} & \eqi{-0.92} & \eqi{-1.75} & \eqi{-1.65 \pm 0.17} \\
\eqi{\delta} & \eqi{+1.65} & \eqi{+1.25} & \eqi{+1.78 \pm 0.26} \\
\eqi{\omega} & \eqi{0} & \eqi{0.125} & \eqi{0.125} (fixed) \\
\eqi{\tau} & \eqi{200} & \eqi{750} & \eqi{2400} (fixed) }

In \firstfig{HR_modified_HR_ohm_total} we present our scaling law's prediction using the exponents as fitted to the data.
We find that the population of the strongest heated field lines follows the equality line between scaling law and our model heating rate very well.
As the prediction is valid only for the coronal loops with a high dissipation of the energy input and we know that most of the coronal field lines are not heated enough to evolve into a EUV-bright loop, we also see here many field lines with a heating rate significantly lower than predicted (below the equality line).

In \tab{exponents} we give the exponents of the \cite{vanBallegooijen:2011} and the \cite{Rappazzo+al:2008} scaling laws (\eqns{Ballegooijen}{Rappazzo}) in comparison to our modified scaling law (\eqn{modified.scaling}) together with the 1-\eqi{\sigma} uncertainty of the fit as estimated by the LM method.

The dependency of the Ohmic heating from the magnetic field strength and field-line length is stronger in our model data than has been predicted by \cite{vanBallegooijen:2011}.
This reflects the influence of geometrical effects due to the curvature of the field in our model, among other effects.
In contrast to this work, the original setup of \cite{vanBallegooijen:2011} used coronal loops that are straight structures and hence depend less on the field-line length \eqi{L}.
A curvature in the magnetic field might amplify the induced coronal currents as compared to a straight loop because the Ohmic heating is proportional to the currents as \eqi{H_{\rm{Ohm}} \sim \vec{j}^2} that have the proportionality \eqi{\vec{j} \sim \vec{\nabla} \times \vec{B}} where the curvature of the magnetic field becomes relevant.

For the magnetic field strength we find a significantly weaker dependency \eqi{\beta} and in our case the driving motions have a stronger impact on the heating than has been predicted by the \cite{Rappazzo+al:2008} scaling law.
Only the fit of the exponent \eqi{\gamma}, representing the dependency on the field-line length, is still consistent within its error as compared to the exponents given in the work of \cite{Rappazzo+al:2008}.

%==============================================================================
\section{Conclusions\label{S:conclusions}}
%==============================================================================
The differences between the prediction of the \cite{vanBallegooijen:2011} scaling law and our model can be explained by a combination of a) different Alfv{\'e}n travel times into the corona, b) different dissipation scales, c) the curvature of the magnetic field in our model setup, and d) a different magnetic energy transport and dissipation mechanism that in our case is more similar to slow magnetic diffusion than to Alfv{\'e}n waves or MHD turbulence.

Even though \cite{Rappazzo+al:2008} use a very similar coronal heating mechanism to the one we implemented in our model (field-line braiding), we still find significantly lower heating rates in our model than the \cite{Rappazzo+al:2008} scaling law would predict.
One possible reason for this difference can be found in the magnetic Reynolds number.
In their comparison of simulations with different magnetic Reynolds numbers, \cite{Rappazzo+al:2008} have shown that a saturation of the heating rates occurs at Reynolds numbers of about 800 and above.
Here our model falls short.
Because we consider a whole active region and each individual coronal loop covers only a fraction of the volume, the magnetic Reynolds number (when considering the length scale of a coronal loop in our model) is significantly lower.
While our whole computational domain covers more grid points (\eqi{1024 \times 1024 \times 256}) than the \cite{Rappazzo+al:2008} models (up to \eqi{512 \times 512 \times 200}), in their model they describe only a single loop structure (straightened
out), and consequently can reach significantly higher Reynolds numbers.

The coronal heating scaling law we fit to our model data indicates a dependency on the magnetic field that lies between the exponents given by \cite{Rappazzo+al:2008} and \cite{vanBallegooijen:2011}.
 We also find that the homogeneous atmospheric background in the MHD-turbulence setup of \cite{Rappazzo+al:2008} may be problematic when estimating the coronal magnetic energy input or braiding of field lines because the Alfv{\'e}n crossing time \eqi{\tau}, which determines their scaling law exponents, varies strongly within our stratified atmosphere.

In a recent study \cite{vanBallegooijen:2014} speculate that the quasi-static behavior of the coronal magnetic field found in models like ours is due to the limited spatial resolution of the photospheric magnetic driving.
They suggest that in the case of increased resolution, the heating might be wave-dominated.
Certainly, with increasing resolution AC heating will also be present, but this does not automatically imply that the DC heating becomes less efficient.
To the contrary, models that employ reduced MHD (allowing for a higher spatial resolution) show that MHD turbulence will lead to a DC-type heating by the build-up of current sheets also for high spatial resolution \citep[e.g.,][]{Rappazzo+al:2008}.
This is backed by the argument that in the coronal part of the atmosphere a field line will adapt quickly to any wave excitation from below because coronal Alfv{\'e}n speeds are at least one order of magnitude faster than within the lower atmosphere (cf. \fig{spectral.index.histogram} and \tab{alpha}).
More realistic full MHD simulations with increased resolution will have to show the relative merits of AC and DC heating based on the photospheric driving.

The temperature scaling laws we tested against our model field lines \citep{Rosner+al:1978,Serio+al:1981} indicate only general trends for the most heated field lines and EUV-emissive loops.
This puts all attempts to set up scaling laws of the coronal heating into their context, namely that these laws are deduced from and are only applicable to a small subset of coronal field lines, namely the ones that are bright in EUV or X-rays.
Non-emissive and weakly heated field lines are often not consistent with scaling laws intended for 1D loop structures.
Some of their plasma parameters need to be determined differently, for example, through self-consistent 3D~MHD simulations.

It is an interesting question whether such observationally motivated scaling laws would benefit from new and much more precise observations of hot coronal loops.
We find that the field lines hosting the hottest plasma in our model roughly follow the old RTV scaling law.
However, with observations one would always miss the cooler coronal field lines because they are not bright in EUV.
On the other hand, self-consistent 3D~MHD simulations provide good data to match all kinds of field lines, as is shown in \sect{fitted.scaling}.

Nonetheless, for EUV-emissive coronal loops and relatively strong heated field lines, we are able to fit a scaling law to our 3D~MHD model data that is consistent with coronal observations \citep[see ][]{Bourdin+al:2013_overview,Bourdin+al:2014_coronal-loops}.
This scaling law includes parameters such as the loop length, the magnetic flux at the footpoints, the photospheric driving motions, the loop density, and the Alfv{\'e}n crossing time in order to roughly predict a coronal Ohmic heat input.

%==============================================================================

\begin{acknowledgements}
This work was supported by the International Max-Planck Research School (IMPRS) on Solar System Physics and was partially funded by the Max-Planck-Princeton Center for Plasma Physics (MPPC).
The results of this research have been achieved using the PRACE Research Infrastructure resource \emph{Curie} based in France at TGCC, as well as \emph{JuRoPA} hosted by the J{\"u}lich Supercomputing Centre in Germany.
Preparatory work has been executed at the Kiepenheuer-Institut f{\"u}r Sonnenphysik in Freiburg, as well as on the bwGRiD facility located at the Universit{\"a}t Freiburg, Germany.
We thank Suguru Kamio for his help finding active region observations.
Hinode is a Japanese mission developed, launched, and operated by ISAS/JAXA, in partnership with NAOJ, NASA, and STFC (UK). Additional operational support is provided by ESA and NSC (Norway).
Finally, we would like to thank the anonymous referee for the valuable suggestions on additional analyses.
\end{acknowledgements}

\bibliography{Literatur}
\bibliographystyle{aa}

\begin{appendix}

%------------------------------------------------------------------------------
\section{Field-line individual analysis of spectral indices\label{A:individual.analysis}}
%------------------------------------------------------------------------------

In \sect{fitted.scaling} we fit the exponents in \eqn{modified.scaling} as independent variables to compare them with other scaling laws.
Strictly speaking, according to the scaling laws derived by \cite{Rappazzo+al:2008} the parameters \eqi{\beta}, \eqi{\delta}, \eqi{\gamma}, and \eqi{\omega} are not independent. Instead, their values are set by one single parameter \eqi{z} defined as
\eql{z = (\alpha + 1) / (\alpha + 2), \label{E:zeta-exponent}}
where \eqi{\alpha} depends only on the Alfv{\'e}n crossing time \eqi{\tau} along each field line (see \tab{alpha}).
We expect variations of the scaling law exponents for each individual field line, depending on the value of \eqi{\tau}.
In the following we check how different these exponents really are for our model data and whether this difference influences our fit.

Because of the density stratification in our model we have a much wider range of Alfv{\'e}n speeds than \cite{Rappazzo+al:2008}.
Thus, we extrapolate the parameter \eqi{\alpha} based on a power law between \eqi{\alpha} and \eqi{c_A} in the range of 50--2000\unit{km/s} given by \cite{Rappazzo+al:2008} and list the values in \tab{alpha}.

\tabl{Alfv{\'e}n speed \eqi{c_A}, the scaling law parameters \eqi{\alpha} and \eqi{z}, and the spectral index \eqi{s}.}{r r l l}{\eqi{c_A}\unit{[km/s]} & \eqi{\alpha} & \eqi{z} & \eqi{s}}{alpha}{
\eqi{1.25} & \eqi{0.01} & \eqi{0.5025} & \eqi{1.671} \\
\eqi{14} & \eqi{0.1} & \eqi{0.5238} & \eqi{1.710} \\
\eqi{50} & \eqi{1/3} & \eqi{0.5714} & \eqi{1.8} \\
\eqi{200} & \eqi{1} & \eqi{0.6667} & \eqi{2} \\
\eqi{310} & \eqi{2} & \eqi{0.75} & \eqi{2.2} \\ % c_A = 309.7
\eqi{400} & \eqi{3} & \eqi{0.8} & \eqi{2.333} \\
\eqi{1000} & \eqi{31/3} & \eqi{0.9189} & \eqi{2.7} \\
\eqi{2000} & \eqi{22} & \eqi{0.9583} & \eqi{2.84} \\
\eqi{8400} & \eqi{100} & \eqi{0.9902} & \eqi{2.961} \\
\eqi{20000} & \eqi{240} & \eqi{0.9959} & \eqi{2.984}}

Despite the wide range of Alfv{\'e}n crossing times (\eqi{c_A} covers over 4 orders of magnitude in \tab{alpha}), the values of \eqi{z} are within a range of only about 0.5--1.0.
Thus, using the mean of \eqi{z=0.75} (corresponding to \eqi{c_A=310\unit{km/s}} or \eqi{\alpha=2}) seems to be a good choice; this value has also been used in \cite{Lionello+al:2013} and \cite{vanWettum+al:2013}.
Now the exponents in \eqn{Rappazzo} are given by
\eqa{\beta &=& 1 + z \nonumber\\
\gamma &=& -1 - z \nonumber\\
\delta &=& 2 - z \nonumber\\
\omega &=& (1 - z) / 2. \nonumber}

%------------------------------------------------------------------------------
\graph{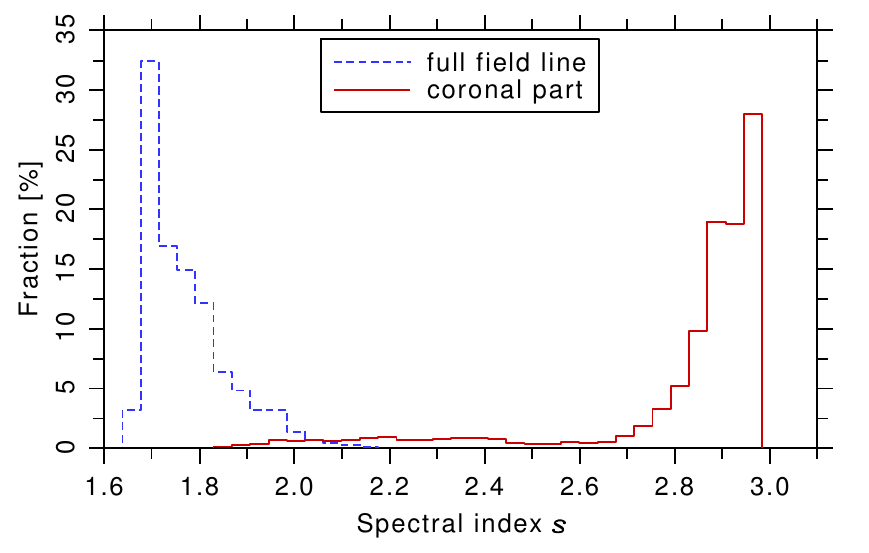}{spectral.index.histogram}{
Histogram (40 bins) of the spectral index \eqi{s} of the magnetic energy conversion along field lines crossing the whole atmosphere (blue dashed) versus only their coronal part (red solid).}
%------------------------------------------------------------------------------

The double-logarithmic slope of the magnetic energy spectrum is set by a turbulent energy conversion cascade along a field line.
This slope is the spectral index \eqi{s} that can be derived as \citep[see Eqn.\,53 in][]{Rappazzo+al:2008}
\eql{s = (3 \alpha + 5) / (\alpha + 3). \label{E:spectral-index}}
In \firstfig{spectral.index.histogram} we show the distribution of the spectral indices as derived from our model field lines.
We find that the spectral index \eqi{s} varies very little for our model field lines, which is connected to the parameter \eqi{\alpha} by \eqn{spectral-index}.
Therefore, it is justified to set \eqi{\alpha} globally constant for all field lines in \sect{Rappazzo.scaling}, even though our choice of \eqi{\alpha = 2} was a bit too high or too low, depending on how one likes to interprete the strongly varying Alfv{\'e}n speed \eqi{c_A} within a stratified atmosphere (see the spectral index of full field lines versus their coronal part in \fig{spectral.index.histogram}).

%------------------------------------------------------------------------------
\graph{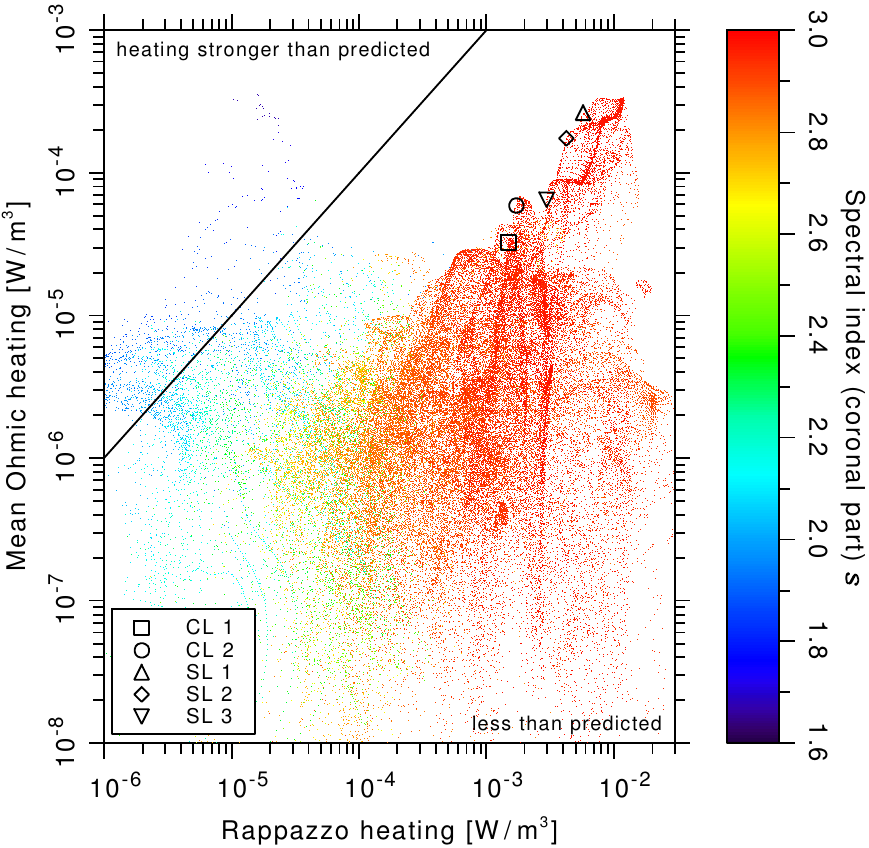}{HR_Rappazzo-alpha_HR_ohm_total}{
Same as \fig{HR_Rappazzo_HR_ohm_total}, but with a field-line individual \eqi{\tau} (and hence~\eqi{z}) parameter. The color-coding represents the spectral index \eqi{s} of the coronal part of each field line, which determines here the exponents of the \cite{Rappazzo+al:2008} scaling law.}
%------------------------------------------------------------------------------

When we do a field-line individual computation of the \cite{Rappazzo+al:2008} scaling law with the actual \eqi{\tau} along each field line, we find in \firstfig{HR_Rappazzo-alpha_HR_ohm_total} that the scatter is greater than in \fig{HR_Rappazzo_HR_ohm_total}.
This is a result of the varying spectral index because the lower \eqi{s} values (blue, on the left) lead to a lower heating rate prediction, while the higher \eqi{s} values (red, on the right) predict an even higher heating rate.
The overestimation of the heating rate for field lines similar to the EUV-bright coronal loops (symbols) increases by roughly one order of magnitude.
Still, the general shape of the distribution of data points remains similar to that in \fig{HR_Rappazzo_HR_ohm_total} because the spectral index \eqi{s} is quite constant within the corona for a large part of our model data (see also the histogram in \fig{spectral.index.histogram}).

In principle, lower \eqi{c_A} values below the corona (which imply lower \eqi{\alpha} and higher \eqi{\tau}) should not be very relevant for the heating rate predicted for the coronal part along the field lines.
Nonetheless, we also checked that lower average \eqi{\alpha} values along the full field lines we find in our model data still result in a excessive heating rate prediction by the \cite{Rappazzo+al:2008} scaling law, very similar to \fig{HR_Rappazzo_HR_ohm_total}, but with slightly increased scatter, for similar reasons to those described above for \fig{HR_Rappazzo-alpha_HR_ohm_total}.
This means that our choice of the \eqi{\alpha} parameter was on the conservative side and that a field-line individual computation does not improve the match between the \cite{Rappazzo+al:2008} scaling law and our model corona.
\end{appendix}

\end{document}

%% file: Journal.inc
\renewcommand*\aap{A\&A}

\renewcommand*\ao{Appl.~Opt.}

\renewcommand*\apjl{ApJL}

\renewcommand*\apjs{ApJS}